\documentclass[aps,preprint,prd]{revtex4}
\usepackage{amssymb}
\usepackage{amsmath}
\usepackage{graphicx}
\usepackage{fancyhdr}
\def\br{\begin{eqnarray}}
\def\er{\end{eqnarray}}
\def\be{\begin{equation}}
\def\ee{\end{equation}}

\def\({\left(}
\def\){\right)}

\newcommand \beq { \begin{eqnarray} }
\newcommand \eeq { \end{eqnarray} }
\newcommand \beqq { \begin{equation} }
\newcommand \eeqq { \end{equation} }
\begin{document}

\title{Combining type I and type II seesaw mechanisms in the minimal 3-3-1 model.}

\author{W. Caetano$^{1}$,  D. Cogollo$^{2}$, C. A. de S. Pires$^{1}$, P. S. Rodrigues da Silva$^{1}$}

\affiliation{\vspace{0.8cm}
\\
 $^{1}$ Departamento de
F\'{\i}sica, Universidade Federal da Para\'\i ba, Caixa Postal 5008, 58051-970,
Jo\~ao Pessoa, PB, Brasi\vspace{0.4cm}
 \\
 $^{2}$ Departamento de
F\'{\i}sica, Universidade Federal de Campina Grande, Caixa Postal 10071, 58109-970,
Campina Grande, Para\'{i}ba, Brazil \vspace{1cm}
 }

\date{\today}

\begin{abstract}
The minimal 3-3-1 model is perturbative until energies around 4-5TeV,  posing a challenge to generate  neutrino masses at eV scale, mainly if one aims to take advantage of the seesaw mechanism. As a means to circumvent this problem we propose a modification of the model such that it accommodates the type I and type II seesaw mechanisms altogether. We show that the conjunction of both mechanisms yield a neutrino mass expression suppressed by  a high power of the cutoff scale, $M^5$, in its denominator. With such a suppression term we naturally obtain neutrino masses at eV scale when $M$ is around few TeV. We also investigate  the size of lepton flavor violation through the process $\mu \rightarrow e\gamma$.
\end{abstract}

%

\maketitle
\section{Introduction}
An interesting class of gauge extensions of the standard model (SM)  that is expected to manifest  at TeV scale is the $SU(3)_C \times SU(3)_L \times U(1)_N$ gauge group for the strong and electroweak interactions, the so-called 3-3-1 model. Among its versions,  the  minimal 3-3-1 model~\cite{vicente} is considered the most interesting one because of its phenomenological aspects based on the presence of doubly charged gauge bileptons, $U^{++}$, and new quarks with exotic electric charges $\frac{5}{3}e$ and  $\frac{4}{3}e$. In its original version three triplets and one sextet of scalars were advocated in order to generate the correct charged fermion mass spectrum~\cite{foot}. Even with such dilated scalar sector the model is not able to generate small neutrino masses.

Recently it was shown that the  minimal 3-3-1 model can be implemented with two triplet of scalars only (reduced version)~\cite{reduced331}. Such short scalar sector is sufficient to engender the spontaneous breaking of the $SU(3)_C \times SU(3)_L \times U(1)_N$ symmetry to the $SU(3)_C \times U(1)_{em}$ one. The main worry here is about the fermion mass spectrum, since two scalar triplets are not enough to generate all the Yukawa interactions necessary to produce mass for all fermions. However, the model has an interesting peculiarity, namely, its pertubative regime requires $\sin^2\theta_W <0.25$.  Translating this in terms of energy, it was pointed out in Ref.~\cite{landaupole} that the pertubative regime of the model is lost around some few TeV. In other words, the highest energy scale where the model  is pertubatively reliable is about 4-5 TeV, and not much further there is a Landau pole on the Weinberg angle, indicating that the model claims for an extension before this. This allows us to make use of effective dimension-5 operators to generate  masses to some  fermions considering this scale as a cutoff~\cite{reduced331}.

However an energy scale around few TeV faces difficulties to generate light neutrinos because some unpleasant effective  operators may  give rise to large neutrino mass terms, unless we assume some fine tuning on the effective Yukawa couplings.  These operators have to be taken into account  when we assume that the lagrangian at tree level contains terms that explicitly violate lepton number. Thus, in order  to avoid such unpleasant effective operators we prohibit  such terms and assume that  Majorana neutrino masses  arise as consequence of spontaneous breaking of the lepton number. 

The proposal of this work is to solve the problem of small neutrino masses in the framework of the reduced 3-3-1 model in a most economic way. For this we extend the scalar and leptonic content of the model in order to implement the type I  and type II seesaw mechanism in a way where lepton number is spontaneously broken. We show that the conjunction of both seesaw mechanisms result in a neutrino mass expression suppressed by the factor $M^5$ which is sufficient to generate neutrino masses at eV scale for $M$ around few TeVs.

\section{The Mechanism}
\label{3}
The leptonic content of the reduced 3-3-1model~\cite{reduced331} is extended by  adding three new singlet right-handed neutrinos,
\begin{eqnarray}
& & f_{aL} = \left (
\begin{array}{l}
\nu_{aL} \,\, e_{aL} \,\, e^{ c}_{aL}
\end{array}
\right )^T\sim(3\,,\,0)\,,\nu_{a_R}\sim(1,0).
\label{leptonsector}
\end{eqnarray}
where $a=1,2,3$,  and the numbers between parentheses refer to the $SU(3)_L$, $U(1)_N$ transformation properties.

The minimal scalar content required to engender the correct spontaneous breaking of the gauge symmetry is composed by only two scalar triplets,
\begin{equation}\label{expansao}
\rho = \left(\begin{array}{c} 
\rho^{+}\\
\rho^{0}\\
\rho^{++}
\end{array}\right) \sim (1,3,1), \quad
\chi = \left(\begin{array}{c} 
\chi^{-}\\
\chi^{--}\\
\chi^{0}
\end{array}\right) \sim (1,3,-1).
\end{equation}
However, for the proposal we have in mind, it is necessary to extend the scalar sector by adding a third scalar triplet and a scalar singlet, namely
\begin{equation}\label{phi}
\phi = \left(\begin{array}{c} 
\phi^{0}\\
\phi_1^{-}\\
\phi_2^{+}
\end{array}\right) \sim (1,3,0)\,\,,\,\,\, \sigma \sim (1,1,0)
\end{equation}
whose mass parameters in the potential must dominate over those of $\rho$ and $\chi$ fields, which is mandatory such that the whole scheme makes sense. 

In order to avoid unpleasant effective operators it is imperative to impose a $Z_4$ symmetry with the following fields transforming as,
\begin{eqnarray}
&&\rho \rightarrow w^3 \rho\,,\,\chi \rightarrow w^3 \chi\,,\,\phi \rightarrow w^2 \phi\,,\,\sigma \rightarrow w^2 \sigma\,,\,\nu_{a_R} \rightarrow w^3 \nu_{a_R}\, , \,u_{a_R} \rightarrow w^3 u_{a_R}\nonumber \\
&&d_{a_R} \rightarrow w^2 d_{a_R}\,,\,Q_1 \rightarrow w Q_1\,,\,J_1 \rightarrow w^2 J_1\,,\,J_i \rightarrow w^3 J_i\,,\,f_{a_L} \rightarrow w f_{a_L},
\label{Z4}
\end{eqnarray}
where $w=e^{i\frac{\pi}{2}}$, $i=2,3$ while $a=1,2,3$.

In view of this, neutrino mass terms can only arise from the symmetric Yukawa interactions,
\begin{equation}\label{yukawaterms}
{\cal  L} =Y \overline{f}_{L}\phi\nu_{R} + \frac{Y^{\prime}}{2}\sigma\overline{\nu}_{R}^c\nu_{R} + H.c.,
\end{equation}
where $Y$ and $Y^{\prime}$ are  $3\times3$ matrices.  Perceive that Lepton number conservation requires that the singlet scalar be a bilepton (carrying two units of lepton number), $L(\sigma)=-2$, and thus lepton number conservation is attached to the $Z_4$ symmetry.

All neutral fields  $\rho^0$,  $\chi^{ 0}$ , $\phi^0$, $\sigma$  are assumed to develop VEV according to
\begin{eqnarray}
\langle\rho^0\rangle=\frac{v_{\rho}}{\sqrt2},\label{veta}\,\,\,\,\,\,
 \langle\chi^{\prime }\rangle=\frac{v_{\chi^{\prime }}}{\sqrt2}\,\,\,\,\,
\langle\phi^0\rangle=\frac{v_{\phi}}{\sqrt2},\,\,\,\,\,\,
\langle\sigma\rangle=\frac{v_{\sigma}}{\sqrt2}.
\label{vrho}
\end{eqnarray}

When $\sigma$ and $\phi$ both develop VEV different from zero, neutrinos develop  Dirac and Majorana  mass terms,
\begin{equation}\label{yukawaterms}
{\cal  L}_{\mbox{mass}} =\frac{Yv_\phi}{\sqrt{2}} \overline{\nu}_{L}\nu_{R} +\frac{1}{2} \frac{Y^{\prime}v_\sigma }{\sqrt{2}}\overline{\nu}_{R}^c\nu_{R} + H.c.
\end{equation}
Note that when this happens  lepton number was broken spontaneously.

On considering the basis $\nu=(\nu_{L}, \nu_{R}^{c})$, we can write the mass terms above in the form,
\begin{equation}\label{neutrinomassterms}
{\cal L}_{\mbox{mass}} =\frac{1}{2}\bar \nu^c M_\nu \nu + H.c.,
\end{equation}
where,
\begin{equation}M_{\nu}= \left(\begin{array}{cc}
0 & M_D\\
M^T_D&  M_R\\
\end{array}\right),
\end{equation}
with $M_D=\frac{Yv_{\phi}}{\sqrt{2}}$ and $M_R=\frac{Y^{\prime}v_{\sigma}}{\sqrt{2}}$. When all the eigenvalues of $M_R$  are larger than all elements of  $ M_D$, we obtain, after diagonalizing this mass matrix, the following expressions for the left and right-handed neutrino masses
\begin{equation} \label{matriz}
m_{\nu_{l}}  \approx M^T_D M_{R}^{-1} M_D, \quad m_{\nu_{R}} \approx M_R.
\end{equation}

This is the canonical type I seesaw mechansim~\cite{typeI}. As usual we consider  $M_R$ diagonal and degenerate. In practical terms we take $M_R=M I$ where $I$ is the identity matrix and  $M=\frac{Y^{\prime}v_{\sigma}}{\sqrt{2}}$. 

Regarding the left-handed (LH) neutrinos, according to Eq.~(\ref{matriz}) the order of magnitude of their masses  is,
\begin{equation} \label{seesawI}
m_{\nu_{l}}  \approx \frac{v^{2}_{\phi}}{M}.
\end{equation}
We recall that there exists a Landau pole in the version of 3-3-1 model we are working with, and this poses a limit to the perturbative realm for this model, forcing us to adopt the maximum value for $M$ to be around $5$~TeV~\cite{landaupole}. Therefore, if we want to have LH neutrinos at the eV scale, we need $v_\phi \approx10^{-2} $~GeV. There is no lower bound on this parameter, however, an upper bound exists, $v^2_\phi + v^2_\rho=(246)^2$~GeV, which arises because  $v_\rho$, as well as $v_\phi$, both contribute to the mass of the standard charged gauge boson $W^{\pm}$. Thus there is no problem in taking $v_\phi$ as small as we wish as long as we keep $v_\rho \approx 246$~GeV.

A suppressed $v_\phi$ can be obtained  through a kind of type II seesaw mechanism as developed in Ref.~\cite{grimus}. Consider the most general scalar potential invariant under the gauge symmetry, $SU(3)_{C} \otimes SU(3)_{L} \otimes U(1)_{N}$  and also the discrete symmetry, $Z_4$,
\begin{eqnarray} \label{potential}
V(\phi, \rho, \chi, \sigma) & = & \mu_{\phi}^{2}\phi^{\dagger}\phi + \mu_{\rho}^{2}\rho^{\dagger}\rho + \mu_{\chi}^{2}\chi^{\dagger}\chi + \mu_{\sigma}^{2}\sigma^{*}\sigma \nonumber\\
&+& \lambda_{1}(\phi^{\dagger}\phi)^{2} + \lambda_{2}(\rho^{\dagger}\rho)^{2} + \lambda_{3}({\chi}^{\dagger}\chi)^{2} +\lambda_{4} (\sigma^{*}\sigma)^{2} \nonumber\\
&+& \lambda_{5}(\phi^{\dagger}\phi)(\rho^{\dagger}\rho) + \lambda_{6}(\phi^{\dagger}\phi)(\chi^{\dagger}\chi) + \lambda_{7}(\rho^{\dagger}\rho)(\chi^{\dagger}\chi) \nonumber\\
&+& \lambda_{8}(\rho^{\dagger}\phi)(\phi^{\dagger}\rho) + \lambda_{9}(\chi^{\dagger}\phi)(\phi^{\dagger}\chi) + \lambda_{10}(\rho^{\dagger}\chi)(\chi^{\dagger}\rho)\nonumber\\
&+& \lambda_{11}(\phi^{\dagger}\phi)(\sigma^{*}\sigma) + \lambda_{12}(\phi^{\dagger}\phi)(\sigma^{*}\sigma) + \lambda_{13}(\chi^{\dagger}\chi)(\sigma^{*}\sigma) \nonumber\\
&-& \frac{f}{\sqrt{2}} \epsilon^{ijk} \phi_{i} \rho_{j}\chi_{k} +H.c.,
\end{eqnarray}
where $f$ is a free parameter with dimension of mass.

Let us assume the following shift in the neutral scalars,
\begin{equation}
\phi^{0}, \rho^{0}, \chi^{0}, \sigma^{0} \rightarrow \frac{1}{\sqrt{2}} (v_{\phi, \rho, \chi, \sigma} + R_{\phi, \rho, \chi, \sigma} + iI_{\phi, \rho, \chi, \sigma}).
\label{scalarshift}
\end{equation}
Requiring the above choice of VEVs, we are going to have the following set of constraint equations,
\begin{eqnarray}
&&v_{\phi} (\mu_{\phi}^{2} + \lambda_{1}v_{\phi}^{2} + \frac{1}{2}\lambda_{5}v_{\rho}^{2} + \frac{1}{2}\lambda_{6}v_{\chi}^{2} + \frac{1}{2}\lambda_{11}v_{\sigma}^{2}) - \frac{1}{2}f v_{\rho}v_{\chi}  = 0,\nonumber \\
&&v_{\rho} (\mu_{\rho}^{2} + \lambda_{2}v_{\rho}^{2} + \frac{1}{2}\lambda_{5}v_{\phi}^{2} + \frac{1}{2}\lambda_{7}v_{\chi}^{2} + \frac{1}{2}\lambda_{12}v_{\sigma}^{2}) - \frac{1}{2}f v_{\phi}v_{\chi}  = 0,\nonumber \\
&&v_{\chi} (\mu_{\chi}^{2} + \lambda_{3}v_{\chi}^{2} + \frac{1}{2}\lambda_{6}v_{\phi}^{2} + \frac{1}{2}\lambda_{7}v_{\rho}^{2} + \frac{1}{2}\lambda_{13}v_{\sigma}^{2}) - \frac{1}{2}f v_{\phi}v_{\rho}  = 0,\nonumber \\
&&\mu_{\sigma}^{2} + \lambda_{4}v_{\sigma}^{2} + \frac{1}{2}\lambda_{11}v_{\phi}^{2} + \frac{1}{2}\lambda_{12}v_{\rho}^{2} + \frac{1}{2}\lambda_{13}v_{\chi}^{2}  = 0.
\label{vevconstraints}
\end{eqnarray}

 From these four relations, the first of them is the only one that matters to us, since it is the one that is going to furnish the desired relation for $v_\phi$. As we have already discussed, the highest energy scale in this model is around $M\approx 5$~TeV if we want to keep on the perturbative realm. It is this scale that we associate to the scalar $\phi$, meaning that  $\mu_\phi \approx M$. Notice that  $\mu_\phi$ is dominant in the parenthesis in the first relation of the Eq.~(\ref{vevconstraints}). As a result, we obtain from it, 
\begin{equation} \label{seesawII}
v_{\phi} = \dfrac{ f v_{\rho} v_{\chi}}{2 M^2},
\end{equation}
which characterizes the type II seesaw mechanism.

On substituting this expression for $v_\phi$ in Eq.~(\ref{matriz}), we obtain the following expression to the LH neutrino mass,
\begin{equation}
m_{\nu_{l}}=\frac{\sqrt{2}}{8}Y^T Y\frac{ f^2 v^2_\rho v^2_\chi }{M^5}.
\label{finalmassmatrix}
\end{equation}
Here comes the main result of this work.  Although $M$, the highest energy scale available for the model, is only some few TeV, according to Eq.~(\ref{finalmassmatrix}), light neutrino masses (at eV scale) may arise thanks to the suppression by the fifth power of $M$ in the denominator. To better appreciate this point, notice that  for reasonable values of the model parameters, $f=1$~GeV, $v_\rho=246$ ~GeV, $v_\chi=10^3$~GeV and $M=5\times10^3$~GeV, we obtain,
\begin{equation}
m_{\nu_{l}}=3.4 Y^T Y\mbox{eV},
\label{magnitude}
\end{equation}
 which falls exactly at the eV scale (keep in mind that the entries in the matrices $Y$ are Yukawa couplings). The correct neutrino masses are obtained by choosing a fair texture for the matrix $Y$.

As a concrete example  we consider $Y$ symmetric and take the following set of possible values for their  elements,
\begin{eqnarray}
&&y_{11} = 0.0181, \quad y_{12}=y_{21} = 0.00350, \quad y_{13}=y_{31} = -0.0276,\nonumber \\
&& y_{22} = -0.0448, \quad y_{23}=y_{32} =  -0.0767, y_{33} = -0.0394.
\label{illustrativeexample}
\end{eqnarray}

With this  we obtain the following texture for the mass matrix $m_{\nu_{l}}$,
\begin{equation}
m_{\nu_{l}} = \left(\begin{array}{ccc} 
0.003745 & 0.00688 & 0.001086\\
0.00688 & 0.02687 & 0.02163\\
0.001086 & 0.02163 & 0.02787
\end{array}\right)~eV.
\end{equation}
The diagonalization of this mass matrix  yields the following masses for LH neutrinos,
$m_1=  5.7 \times 10^{-5}$ , $m_2= 8.7 \times 10^{-3}$ and $m_3= 5.0 \times 10^{-2}$ which provides the following neutrino mass squared differences:$$ \Delta m^{2}_{21} \equiv m^{2}_{2} - m^{2}_{1} = 7.6 \times 10^{-5} eV^{2}, \quad \Delta m^{2}_{31} \equiv m^{2}_{3} - m^{2}_{1} =  2.5 \times 10^{-3} eV^{2}.
$$

The $U_{PMNS}$ mixing matrix that diagonalizes $m_{\nu_L}$ above is,
\begin{equation}
U_{PMNS} = \left(\begin{array}{ccc} 
0.803 & 0.583 & 0.122\\
-0.485 & 0.521 & 0.702\\
0.346 & -0.623 & 0.702
\end{array}\right),
\end{equation}
which implies  the following mixing angles $\theta_{12} = 36^o$, $\theta_{23} = 45^o$ and e $\theta_{13} = 7^o$. 

Thus, such mixing angles together with the above neutrino mass squared differences, explain both the solar~\cite{solar}and atmospheric~\cite{atm} neutrino oscillations and the recent experimental results concerning the angle $\theta_{13}$~\cite{exp.theta13} 

The mixing among LH and RH neutrinos is given by $V=M_D M_R^{-1}$.  In this way the LH neutrinos composing the leptonic charged current  interactions of the SM are a superposition of six neutrinos, $(\hat \nu_L\,\,,\,\,N_L)$, which in first order in $M_D M_R^{-1}$ can be written as,
\begin{equation}
 \nu_L \approx U_{PMNS} \hat \nu_L + V N_L.
\label{neutrinomixing}
\end{equation}
Then the charged current in the physical basis is,
\begin{eqnarray}
{\cal L}_{CC}&=&-\frac{g}{\sqrt{2}}\bar f_{aL} \gamma^\mu  \nu_{aL} W^-_\mu +H.c.
\nonumber\\
&\approx &-\frac{g}{\sqrt{2}}\bar f_{aL} \gamma^\mu\left(U_{PMNS}\hat \nu_{aL} +V N_{aL} \right)W^-_\mu +H.c.
\label{CC}
\end{eqnarray}
As we are assuming that the  RH neutrinos  have masses around few TeV, it becomes important to check if their contributions to lepton flavor violation (LFV) processes respect  current bounds. The most constraining LFV process is $\mu \rightarrow e\gamma$ whose current upper bound on its branching ratio is $BR(\mu \rightarrow e \gamma)< 4.9\times 10^{-11}$~\cite{pdg}. By taking the above considerations on neutrino mixing, we can compute this branching ratio for our model.
 
The expression for the  branching ratio for the process $\mu \rightarrow e\gamma$ is given by Ref.~\cite{BR},
\begin{equation}
BR(\mu \rightarrow e \gamma)\approx \frac{\alpha^3_W \sin^2(\theta_W) m^5_{\mu}}{256 \pi^2 m^4_W \Gamma_\mu}\times |(VV^T)_{21}I(\frac{m^2_{\nu_R}}{m^2_W})|^2,
\label{BR}
\end{equation}
where
\begin{equation}
I(x)=-\frac{2x^3+5x^2-x}{4(1-x)^3}-\frac{3x^3lnx}{2(1-x)^4}.
\label{argument}
\end{equation}
In the branching ratio above $\alpha_W= \frac{g^2}{4 \pi}$ with $g$ being the weak coupling constant, $\theta_W$ is the Weinberg's angle, $m_\mu$ is the mass of the muon, $m_W$ is the mass of the $W^{\pm}$, $\Gamma_\mu$ is the total muon decay width. The precise values of these parameters can be extracted from Ref.~\cite{pdg}.

With $M_D$ and $M_R$ taken from the illustrative example above we obtain the following values for the branching ratio, 
\begin{equation}
BR(\mu \rightarrow e \gamma)\approx 1.9\times 10^{-26},
\label{BRvalue}
\end{equation}
which respects the current upper bounds but, unfortunately,  is too small escaping the sensitivity of  future neutrino experiments which will probe  branching ratios of order up to $10^{-18}$~\cite{future}. In other worlds, although the seesaw mechanism developed here involves right-handed neutrinos with mass around few TeV, their mixing with the standard neutrinos are very suppressed staying, as in the canonical case, far from  being detected in the present or future neutrino experiments. This result should remain for whatever set of values we take for the Yukawa couplings, $Y$,  once they would be not far from the order of magnitude we used in the illustrative example above.

\section{Conclusions}
 The highest energy scale where the minimal 3-3-1 model allows for reasonable perturbative computations is about 4-5 TeV, due to the existence of a Landau pole evidenced by the running of the Weinberg's angle~\cite{landaupole}. This peculiar property of this gauge model turns out to be interesting in the sense that we can make use of this fact  to generate masses for  fermions through effective operators. This allows a remarkable reduction of the number of degree of freedom on the scalar sector of the model, which was previously thought to contain at least three triplets and one or two sextets. It was recently observed in Ref.~\cite{reduced331} that two triplet of scalars are sufficient to engender the correct spontaneous breaking of the $SU(3)_C \times SU(3)_L \times U(1)_N$ symmetry to the $ \rightarrow SU(3)_C \times U(1)_{em}$ and generate masses for all the fermions. 
Some fine tuning in the effective Yukawa couplings is still manadatory in this scheme though, mainly for the neutrinos.
In order to completely avoid such fine tuning concerning neutrino's masses and profit from the elegant seesaw mechanism in this model, we noticed that the implied low cut off energy poses a big challenge to such a procedure. This is because, as it is well known, neutrino masses around eV scale demand a high energy scale when  explanation is via seesaw mechanism. 

In this work we have slightly modified the reduced 3-3-1 model from its original form presented in Ref.~\cite{reduced331} in order to circumvent the above mentioned difficulty in generating naturally small neutrino masses. We performed this change by adding three singlet RH neutrinos, as well as a triplet and a singlet scalars. Such modification was chosen in an appropriate way so that we could build the type I and type II seesaw mechanisms and combine them to yield neutrino masses suppressed by a high-scale $M^5$ in its denominator and, in this way, we got neutrino masses at eV scale naturally for $M$ around few TeV. Regarding some  characteristic signature of the mechanism developed here, unfortunately, and similar to the canonical case, it can leave no track in the present or planned experiments. This is so because, even with RH neutrinos having masses around few TeV, the mixing among LH and RH neutrinos get  very suppressed due to the type II seesaw mechanism on the VEV responsible by the Dirac mass $M_D$.  

\acknowledgments
This work was supported by Conselho Nacional de Pesquisa e
Desenvolvimento Cient\'{i}fico- CNPq ( CASP and PSRS) and Coordena\c c\~ao de Aperfei\c coamento de Pessoal de N\'{i}vel Superior - CAPES (WC).
\section*{Appendix A}

In this appendix we check if the VEVs of the scalar sector are sufficient for generating the correct masses of the quarks. For this note that the quark sector is the original one where~\cite{vicente},
\begin{eqnarray}
&&
\begin{array}{c}
 Q_{1L}=\left(\begin{array}{cc}
 u_1 \\ d_1 \\ J_1
\end{array}\right)_L \sim (3,3,+\frac{2}{3})
\end{array}\,\,,\,\,
\begin{array}{c}
Q_{iL}=\left(\begin{array}{cc}
 d_i \\ -u_i \\ J_i
\end{array}\right)_L \sim (3,3^*,-\frac{1}{3}),
\end{array}
\nonumber \\
&&
\begin{array}{ccc}
u_{iR} \sim(3,1,+\frac{2}{3}); & d_{iR}
\sim(3,1,-\frac{1}{3}); & J_{iR}
\sim(3,1,-\frac{4}{3}),
\end{array}
\nonumber \\
&&
\begin{array}{ccc}
u_{1R} \sim(3,1,+\frac{2}{3}); & d_{1R}
\sim(3,1,-\frac{1}{3}); & J_{1R}
\sim(3,1,+\frac{5}{3}),
\end{array}
\end{eqnarray}
\label{quarkcontent}
with $i=2,3$.

The Yukawa interactions among scalars and   quarks are composed by the following terms,
\begin{eqnarray}
&& \lambda_{11}^J \bar Q_{1_L}\chi J_{1R}+ \lambda^J_{ij_L} \bar Q_{iL}\chi^* J_{j_R}+\nonumber \\
&&\lambda^d_{1a}\bar Q_{1_L} \rho d_{a_R}+ \lambda^d_{ia_L}\bar Q_{iL}\phi^* d_{a_R}+\nonumber \\
&&\lambda^u_{1a}\bar Q_{1_L} \phi u_{a_R}+ \lambda^u_{ia_L}\bar Q_{iL}\rho^* u_{a_R}+H.c.\,,
\end{eqnarray}
where $i,j=2,3$ and $a=1,2,3$.

We should be concerned whether the spectrum of scalars used here, where one of them develops a small VEV, $v_\phi$,  is able to provide the correct values for the quark masses. Fortunately, the up-type and down-type quark's masses both have origin in Yukawa interactions involving the triplet $\phi$ as well as the triplet $\rho$. This is sufficient to guarantee that a small $v_\phi$ together with a standard $v_\rho$  produce the correct masses for all quarks.

Let us consider the set of VEVs taken in the body of the manuscript. For the up-type quarks, the Yukawa interactions yield the following mass matrix in the basis $(u_1\,, u_2\,, u_3)$,
\begin{equation}
M^{u} = \frac{1}{\sqrt{2}} \left(\begin{array}{ccc} 
\lambda^{u}_{11} v_{\phi} & \lambda^{u}_{12} v_{\phi} & \lambda^{u}_{13} v_{\phi}\\
-\lambda^{u}_{21} v_{\rho} & -\lambda^{u}_{22} v_{\rho} & -\lambda^{u}_{23} v_{\rho}\\
-\lambda^{u}_{31} v_{\rho} & -\lambda^{u}_{32} v_{\rho} & -\lambda^{u}_{33} v_{\rho}
\end{array}\right).
\end{equation}
For this set of Yukawa couplings,
\begin{eqnarray}
&&\lambda^{u}_{11} = 0.30, \quad \lambda^{u}_{12} = 0.02, \quad \lambda^{u}_{13} = 0.04;\nonumber\\
&& \lambda^{u}_{21} = -0.04, \quad \lambda^{u}_{22} = -0.005, \quad \lambda^{u}_{23} = -0.05;\nonumber\\
&& \lambda^{u}_{31} = 0.03, \quad \lambda^{u}_{32} = 0.04, \quad \lambda^{u}_{33} = -0.99,
\label{yukawaUP}
\end{eqnarray}
we obtain,

$$ m_{u} \approx 3.0 MeV \quad  m_{c} \approx 1.22 GeV \quad m_{t} \approx 171.8 GeV .$$

Now, for the  down-type  quarks, the Yukawa interactions yield the following mass matrix in the basis $(d_1\,, d_2\,, d_3)$,
\begin{equation}
M^{d} = \frac{1}{\sqrt{2}} \left(\begin{array}{ccc} 
\lambda^{d}_{11} v_{\rho} & \lambda^{d}_{12} v_{\rho} & \lambda^{d}_{13} v_{\rho}\\
\lambda^{d}_{21} v_{\phi} & \lambda^{d}_{22} v_{\phi} & \lambda^{d}_{23} v_{\phi}\\
\lambda^{d}_{31} v_{\phi} & \lambda^{d}_{32} v_{\phi} & \lambda^{d}_{33} v_{\phi}
\end{array}\right),
\end{equation}
for the following set of Yukawa couplings,
\begin{eqnarray}
\lambda^{d}_{11} = 0.025; \quad \lambda^{d}_{12} = 0.2; \quad \lambda^{d}_{13} = 0.4;\nonumber\\
\lambda^{d}_{21} = -0.4; \quad \lambda^{d}_{22} = -0.03; \quad \lambda^{d}_{23} = -0.7;\nonumber\\
\lambda^{d}_{31} = -0.173; \quad \lambda^{d}_{32} = -0.12; \quad \lambda^{d}_{33} = 0.015,
\end{eqnarray}
the down-type quark's masses will be given by,

$$ m_{d} \approx 4.9 MeV \quad  m_{c} \approx 101 MeV \quad m_{t} \approx 4.24 GeV. $$

As we do not have a scalar sextet, the mass of the charged leptons arise from the following dimension-5 effective operator that conserves lepton number,
\begin{equation}
\frac{Y^{\prime\prime}}{\Lambda}\left(\overline{L^c_L}\rho^*\right)\left(\chi^\dagger L_L \right) + h.c.
\label{chargedleptonmasses}
\end{equation}
This effective operator gives the following mass term for the charged leptons $m_l\approx \frac{1}{2}Y^{\prime\prime}v_\rho$, which easily reproduces the charged lepton masses as long as the Yukawa couplings $Y^{\prime\prime}$ are similar to the standard model ones.

\section*{Appendix B}

In this appendix  we check if the potential given in Eq.~(\ref{potential}) is stable for the set of VEVs we used in this work. For this we consider the shift in the neutral scalars given in Eq.~(\ref{scalarshift}) and the corresponding constraint equations given in Eq.~(\ref{vevconstraints}).

After imposing the constraint equations, the squared mass matrix for the doubly charged scalar fields in the 
$(\chi^{++},\rho^{++})$ basis has the following form,
\begin{equation} \label{doubly}
M^{2}_{H_{++}} = \left(\begin{array}{cc} 
\frac{fv_{\phi}v_{\chi}}{2v_{\rho}} + \frac{\lambda_{10}v_{\chi}^{2}}{2} & \frac{fv_{\phi}}{2} + \frac{\lambda_{10}v_{\rho}v{\chi}}{2}\\
\frac{fv_{\phi}}{2} + \frac{\lambda_{10}v_{\rho}v_{\chi}}{2} & \frac{fv_{\phi}v_{\rho}}{2v_{\chi}} + \frac{\lambda_{10}v_{\rho}^{2}}{2}
\end{array}\right),
\end{equation}
The eigenstates of the matrix in Eq.~(\ref{doubly}) will be denoted as $h_{1,2}^{++}$. The diagonalization of $M^{2}_{H_{++}}$ gives the following mass spectrum,
\begin{equation}
m^{2}_{h_{1}^{++}}=0,
\end{equation}
\begin{equation}
m^{2}_{h_{2}^{++}}= \dfrac{1}{2}\left( \frac{f v_{\phi}v_{\chi}}{v_{\rho}} + \frac{f v_{\phi}v_{\rho}}{v_{\chi}} +\lambda_{10}(v_{\rho}^{2} + v_{\chi}^{2}) \right).
\label{DCmass}
\end{equation}

We see that to have a positive  $m^2_{h_{2}^{++}}$, it requires $\lambda_{10} >0$.

For the simply charged scalars we obtain the following squared mass matrix in the basis 
$(\phi^{+}_2,\chi^{+},\phi^{+}_1,\rho^{+})$,
\begin{equation} \label{simply}
M^{2}_{H_{+}} = \left(\begin{array}{cccc} 
\frac{ f v_{\rho} v_{\chi}}{2 v_{\phi}} + \frac{\lambda_{9}v_{\chi}^2}{2} & \frac{ f v_{\rho}}{2} + \frac{\lambda_{9} v_{\phi}v_{\chi}}{2} & 0 & 0\\

\frac{ f v_{\rho}}{2} + \frac{\lambda_{9} v_{\phi}v_{\chi}}{2} &\frac{ f v_{\phi} v_{\rho}}{2 v_{\chi}} + \frac{\lambda_{9}v_{\phi}^2}{2} & 0 & 0\\

0 & 0 & \frac{ f v_{\rho} v_{\chi}}{2 v_{\phi}} + \frac{\lambda_{8}v_{\rho}^2}{2} & \frac{ f v_{\chi}}{2} + \frac{\lambda_{8} v_{\phi}v_{\rho}}{2} \\

0 & 0 & \frac{ f v_{\chi}}{2} + \frac{\lambda_{8} v_{\phi}v_{\rho}}{2} & \frac{ f v_{\phi} v_{\chi}}{2 v_{\rho}} + \frac{\lambda_{8}v_{\phi}^2}{2}
\end{array}\right),
\end{equation}

We denote the physical eigenstates as $h_{1,2,3,4}^{+}$.  The diagonalization of $M^{2}_{H_{+}}$ gives  the following mass spectrum,
\begin{eqnarray}
&& m^{2}_{h_{1}^{+}}=0,\,\,\, m^{2}_{h_{2}^{+}}=  \dfrac{1}{2} f v_{\rho} \left( \frac{v_{\chi}}{v_{\phi}} + \frac{v_{\phi}}{v_{\chi}}\right) + \dfrac{\lambda_{9}}{2}(v_{\rho}^{2} + v_{\chi}^{2}),\nonumber \\
&& m^{2}_{h_{3}^{+}}=0,\,\,\,m^{2}_{h_{4}^{+}}=\dfrac{1}{2} f v_{\chi} \left(\frac{v_{\phi}}{v_{\rho}} + \frac{v_{\rho}}{v_{\phi}}\right) + \dfrac{\lambda_{8}}{2}(v_{\rho}^{2} + v_{\chi}^{2}). 
\label{simplycharged}
\end{eqnarray}
The two Goldstone bosons are eaten by the  gauge bosons $W^{\pm}$ and $V^{\pm}$. We see that to have positive  $m^2_{h_{2}^{+}}$ as well as $m^2_{h_{4}^{+}}$, it requires $\lambda_8,\lambda_9>0$.

For the CP-odd neutral scalars, on considering the basis $(I_{\phi^{0}}, I_{\rho^{0}}, I_{\chi^{0}}, I_{\sigma^{0}})$ we obtain the following mass matrix,
\begin{equation} \label{odd}
M^{2}_{I^{0}} = \dfrac{f}{4}
\left(\begin{array}{cccc} 
\dfrac{v_{\rho} v_{\chi}}{v_{\phi}} & v_{\chi}  & v_{\rho} & 0 \\
v_{\chi} & \dfrac{v_{\phi} v_{\chi}}{v_{\rho}}  & v_{\phi} & 0 \\
v_{\rho} & v_{\phi} & \dfrac{v_{\phi} v_{\rho}}{v_{\chi}} & 0 \\
0 & 0 & 0 & 0
\end{array}\right) .
\end{equation}

We denote the  eigenstates of the matrix in Eq. $(\ref{odd})$ as  $g_{1,2,3,4}^{0}$. The CP-odd mass spectrum is, 
\begin{equation}\label{cpoddzeromass}
m^{2}_{g_{1,2,4}^{0}}=0,\nonumber
\end{equation}
and
\begin{equation}\label{cpoddmass}
m^{2}_{g_{3}^{0}} =  \dfrac{1}{2}\left( \frac{f v_{\rho}v_{\chi}}{v_{\phi}} + \frac{f v_{\phi}v_{\chi}}{v_{\rho}} 
+ \frac{f v_{\phi}v_{\rho}}{v_{\chi}} \right).
\end{equation}
Note  that $m_{g_{3}^{0}} $ is positive.

$g^0_{1,2}$ are Goldstone bosons eaten by the neutral gauge bosons $Z^0$ and $Z^{\prime}$. $g^0_4$  is a Majoron.  This Majoron is a consequence of the   spontaneous breaking of the lepton number. Note that $g^0_4$ decoupled from the other CP-odd scalars which means it is a singlet for the standard interactions, consequently it is a safe Majoron. 

The squared mass matrix  for the CP-even neutral components of the scalar sector  in the basis $(R_{\phi^{0}}, R_{\rho^{0}}, R_{\chi^{0}}, R_{\sigma^{0}})$  has the following form,
\begin{equation} \label{even}
M^{2}_{R^{0}} = \left(\begin{array}{cccc} 
\dfrac{ f v_{\rho} v_{\chi}} {4 v_{\phi}} + \lambda_{1}v_{\phi}^{2}
& \dfrac{\lambda_{5}v_{\phi} v_{\rho}}{4} - \dfrac{ f v_{\chi}}{4}  
& \dfrac{\lambda_{6}v_{\phi} v_{\chi}}{4} - \dfrac{ f v_{\rho}}{4}
& \dfrac{\lambda_{11}v_{\phi} v_{s}}{2}\\

\dfrac{\lambda_{4}v_{\phi} v_{\rho}}{4} -\dfrac{ f v_{\chi}}{4} & \dfrac{ f v_{\phi} v_{\chi}}{4 v_{\rho}} + \lambda_{2}v_{\rho}^{2} & \dfrac{\lambda_{6}v_{\rho} v_{\chi}}{4} - \dfrac{ f v_{\phi}}{4} & \dfrac{\lambda_{12}v_{\rho} v_{s}}{2}\\

\dfrac{\lambda_{5}v_{\phi} v_{\chi}}{4} - \dfrac{ f v_{\rho}}{4} & \dfrac{\lambda_{6}v_{\rho} v_{\chi}}{4} - \dfrac{ f v_{\phi}}{4} & \dfrac{ f v_{\phi} v_{\rho}}{4 v_{\chi}} + \lambda_{3}v_{\chi}^{2} & \dfrac{\lambda_{13}v_{\chi} v_{s}}{2}\\

\dfrac{\lambda_{11}v_{\phi} v_{s}}{2} & \dfrac{\lambda_{12}v_{\rho} v_{s}}{2} & \dfrac{\lambda_{13}v_{\chi} v_{s}}{2} & \lambda_{4}v_{s}^{2}

\end{array}\right) 
\end{equation}

We denote the physical eigenstates  as $h_{1,2,3,4}^{0}$. The
$DetM^{2}_{R^{0}} \neq 0$  guarantees that the CP-even sector has no Goldstone bosons. The diagonalization of this matrix is not trivial, but it is straightforward to see that for reasonable values of the $\lambda$s involved in it we are going to have positive masses as required by the stabilization of the potential. Just to have an idea of the values of the scalars' masses, we present in  TABLE I a set of values for the $\lambda$s and the corresponding mass values for the scalars in TABLE II.

\begin{table}
\begin{tabular}{|c|c|}
\hline
Parameter & Value \\
\hline \hline
$\lambda_{1},\lambda_{3-7},\lambda_{11-13}$  & 0.1 \\ \hline
$\lambda_{2}$ & 0.29  \\ \hline
$\lambda_{(8-10)}$ & [0.1 - 0.9] \\ \hline
f & 1 GeV  \\ \hline
$v_{\phi}$ & $0.0246$ GeV \\ \hline
$v_{\rho}$ & $246$ GeV  \\ \hline
$v_{\chi}$ & 1 TeV  \\ \hline
$ M$ & 5 TeV \\ \hline
\end{tabular}
\caption{A possible set of values for the parameters involved in the scalar sector.}
\end{table}
\begin{table}
\begin{tabular}{|c|c|}
\hline
Scalar Spectrum & Mass (GeV) \\
\hline \hline
$m_{h_{1}^{++}}$ & 0  \\ \hline
$m_{h_{2}^{++}}$ & [230.2 - 690.8] for $\lambda_{10} = [0.1 - 0.9]$  \\ \hline
$m_{h_{1,3}^{+}}$ & 0 \\ \hline
$m_{h_{2}^{+}}$ & [2247.2 - 2236.7] for $\lambda_{9} = [0.1 - 0.9]$ \\ \hline
$m_{h_{4}^{+}}$ & [2343.5 - 2242.1] for $\lambda_{8} = [0.1 - 0.9]$ \\ \hline
$m_{g_{1,2}^{0}}$ & 0 \\ \hline
$m_{g_{3}^{0}}$ & 3535.5 \\ \hline
$m_{g_{4}^{0}}$ & 0 \\ \hline
$m_{h_{1}^{0}}$ & 2236.07 \\ \hline
$m_{h_{2}^{0}}$ & 1589.77 \\ \hline
$m_{h_{3}^{0}}$ & 273.56 \\ \hline
$m_{h_{4}^{0}}$ & 125.29 \\ \hline
\end{tabular}
\caption{Possible values for the scalars masses according to the parameters in  TABLE I.}
\end{table}
Notice that the neutral scalar, $h^0_4$, has a mass of $125$~GeV for the chosen values of the free parameters given in TABLE I. Its eigenvector is given by  $ h_{4}^{0} \approx \quad 0.98 \rho^0 + 0.10 \chi^0$. In other words, it is basically the $\rho^0$ component of the triplet $\rho$. This value for the mass of $h^0_4 $ is interesting because of recent hints about the Higgs mass in LHC and Tevatron~\cite{LHCTev}.


\end{document}